\documentclass[preprint,epsfig,floats,aps]{revtex4}
\usepackage{epsfig}

\begin{document}

\title{Electron-Positron Annihilation into Hadron-Antihadron Pairs}
\author{Y. \ Yan \footnotemark[1], C. \ Kobdaj \footnotemark[1],
P. \ Suebka \footnotemark[1], Y. \ M. \ Zheng \footnotemark[1],
Amand \ Faessler \footnotemark[2], Th. \ Gutsche \footnotemark[2],
V. \  E. \ Lyubovitskij \footnotemark[2]\footnotemark[3]
\vspace*{0.4\baselineskip}}
\affiliation{\footnotemark[1]
School of Physics, Suranaree University of Technology, \\
Nakhon Ratchasima 30000, Thailand \\
\footnotemark[2]
Institut f\"ur Theoretische Physik, Universit\"at T\"ubingen,
Auf der Morgenstelle 14, D-72076 T\"ubingen, Germany\\
\footnotemark[3]
On leave of absence from Department of Physics, \\
Tomsk State University, 634050 Tomsk, Russia
\vspace*{0.4\baselineskip}}

\begin{abstract}
The reactions $e^+e^-\rightarrow\pi^+\pi^-$ and
$e^+e^-\rightarrow  \overline NN$ with $N=p, n$ are studied
in a non-perturbative quark model. The
work suggests that the two-step process, in which the primary
$\overline qq$ pair forms first a vector meson which in turn
decays into a hadron pair, is dominant over the one-step process
in which the primary $\overline qq$ pair is directly dressed by
additional $\overline qq$ pairs to form a hadron pair. To
reproduce the experimental data of the reaction
$e^+e^-\rightarrow\overline nn$ and $\overline pp$ a $D$-wave
$\omega$-like vector meson with a mass of around 2 GeV has to be
introduced.
\end{abstract}

\maketitle

\section{Introduction}
\begin{figure}
\centerline{\hspace{-0.5in}
\epsfig{file=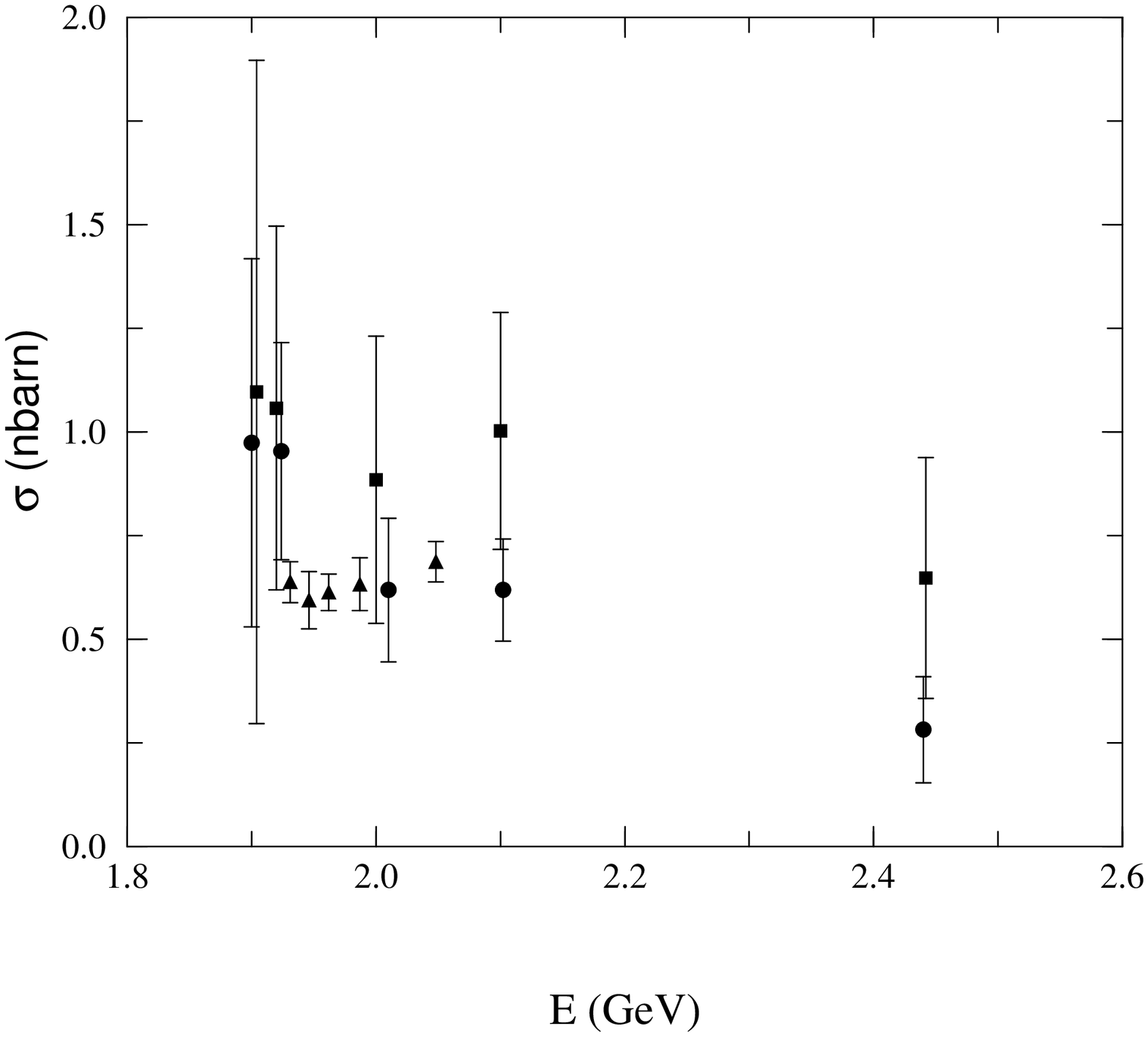,width=3.5in,height=4.5in,angle=270}}
\vspace{0.2in} \caption{Comparison of the cross sections for
$e^+e^-\rightarrow\overline pp$ and $e^+e^-\rightarrow\overline
nn$ at the $\overline NN$ threshold region.
The solid circles~\cite{eebar2} and triangles \cite{eebar3} are
for the reaction
$e^+e^-\rightarrow\overline pp$ while the squares~\cite{eebar1}
are for $e^+e^-\rightarrow\overline nn$.} \label{eeNN-exp}
\end{figure}

Experimental data on the reaction $e^+e^-\rightarrow\overline nn$
from the FENICE collaboration~\cite{eebar1}, earlier data on the
reaction $e^+e^-\rightarrow\overline pp$ from
the FENICE and DM2 collaborations~\cite{eebar2} and also
data collected at the LEAR antiproton ring at CERN on the
time-reversed reaction $\overline pp\rightarrow
e^+e^-$~\cite{eebar3} which are summarized in FIG.\ref{eeNN-exp}
indicate a ratio
$\sigma(e^+e^-\rightarrow\overline nn)/
\sigma(e^+e^-\rightarrow\overline pp) > 1$ at energies around
the $\overline NN$ threshold with $E_{c.m.}\sim 2$ GeV. Averaging
over the available data on both the direct and time-reversed
reactions, one finds~\cite{Ellis}
\begin{eqnarray}\label{ratio}
\frac{\sigma(e^+e^-\rightarrow\overline
pp)}{\sigma(e^+e^-\rightarrow\overline nn)} =0.66^{+0.16}_{-0.11}.
\end{eqnarray}
\par
In a naive perturbative description of $e^+e^-$ annihilation into
baryons the virtual time-like photon first decays into a
$\overline qq$ pair, then the $\overline qq$ pair is dressed by
two additional quark-antiquark pairs excited out of the vacuum to
form a baryon pair. The dressing process does not distinguish
between $u$ and $d$ quarks at high momentum transfers since in the
description of perturbative QCD the gluon couplings are flavor
blind. In the conventional perturbative picture the only
difference between the proton and neutron production arises from
the different electric charges of the primary $\overline qq$
pairs. One expects to get
\begin{eqnarray}
\frac{\sigma(e^+e^-\rightarrow\overline
pp)}{\sigma(e^+e^-\rightarrow\overline nn)} > 1
\end{eqnarray}
at large momentum transfers where the $u$ quark contribution
dominates in the proton and the $d$ quark in the neutron.
\par
The reaction $e^+e^-\rightarrow\overline NN$ at energies around
the $\overline NN$ threshold is highly nonperturbative, hence the
problem must be tackled in a nonperturbative manner. In this work
we model the reactions by the nonperturbative $^3P_0$ quark
dynamics which describes quark-antiquark annihilation and
creation. It was shown that the $^3P_0$ approach is
phenomenologically successful in the description of hadronic
couplings~\cite{3P0,tueb1,dgmf,muhn,yanatom}.

The reaction $e^+e^-\rightarrow\overline NN$ may arise from two
different processes: (1) the primary $\overline qq$ pair is
dressed directly by two additional quark-antiquark pairs created
out of the vacuum to form a baryon pair; and (2) the primary
$\overline qq$ pair forms a virtual vector meson first, then the
virtual vector meson decays into a baryon pair. We expect that the
second process is dominant over the first because of the
considerable success of the vector dominance model. However, it is
difficult to extract a solid conclusion by studying the reaction
itself since there are only very limited experimental data
available and the effective strength of the quark-antiquark vertex
may vary largely from one process to another. We therefore study
first a much simpler process, the reaction
$e^+e^-\rightarrow\pi^+\pi^-$, where a large number of high
quality data are available. The work is arranged as follows: In
Sec. II we study the reaction $e^+e^-\rightarrow\pi^+\pi^-$ with
the parameters determined in the reactions $\rho^0\rightarrow e^+e^-$
and $\rho^0\rightarrow\pi^+\pi^-$. The reaction
$e^+e^-\rightarrow\overline NN$ is studied in Sec. III in the
two-step process described above. We give our conclusions in Sec. IV.
In Appendices A and B we discuss the calculations of
the transition amplitudes $\rho^0 \to \pi^+\pi^-$ and
$V \to \overline NN$ in the $^3P_0$ model.

\section{Reaction $e^+e^- \rightarrow \pi^+\pi^-$}

\begin{figure}
\centerline{\hspace{-0.5in}
\epsfig{file=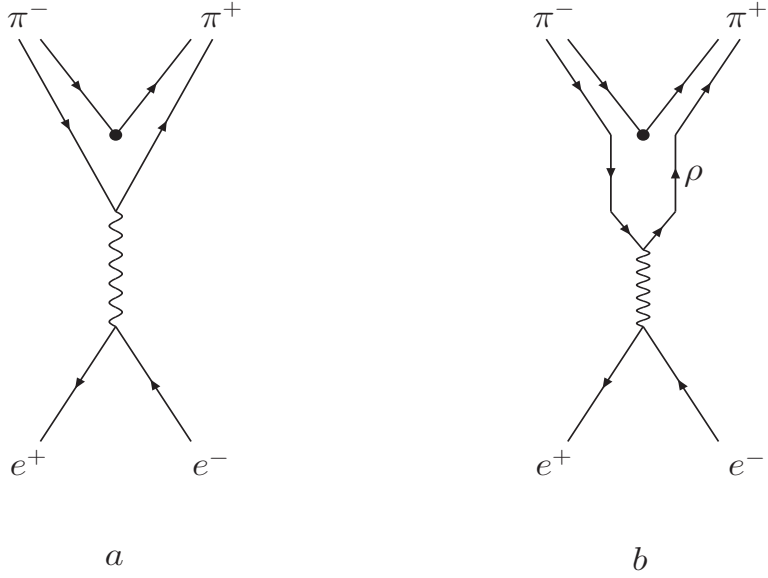,width=4.in,height=3.0in,angle=0}}
\vspace{0.2in} \caption{The reaction $e^+e^-\rightarrow\pi^+\pi^-$
in the one-step process (a), and the two-step process (b).}
\label{eepipi-fig}
\end{figure}

The reaction $e^+e^-\rightarrow\pi^+\pi^-$ may arise in the
valence quark dominated picture from the following process:
the $e^+e^-$ pair annihilates into a virtual time-like photon,
the virtual photon decays into a $\overline qq$ pair, and finally
the $\overline qq$ pair is dressed by an additional
quark-antiquark pair created out of the vacuum to form a meson
pair, as shown in FIG.\ref{eepipi-fig}a. The transition amplitude
is expressed formally as
\begin{eqnarray}\label{onestep}
T_1=\langle\pi^+\pi^-|V_{\overline qq}|\overline qq\rangle
\langle\overline qq|G|\overline qq\rangle\langle\overline
qq|T|e^+e^-\rangle
\end{eqnarray}
where $\langle\overline qq|T|e^+e^-\rangle$ is simply the
transition amplitude of $e^+e^-$ to a primary quark pair,
$\langle\overline qq|G|\overline qq\rangle$ is the Green function
describing the propagation of the intermediate $\overline qq$
state and $\langle\pi^+\pi^-|V_{\overline qq}|\overline qq\rangle$
denotes the amplitude of the process of a $\overline qq$ pair to a
$\pi^+\pi^-$ pair. $V_{\overline qq}$ is the effective vertex for
creation and destruction of a quark-antiquark pair in quark
models, which is identified in the context of the $^3P_0$
quark-antiquark dynamics.
At an energy scale of about 1 GeV
the intermediate quark-antiquark state can be assumed to be saturated
by the $\rho^0(770)$ resonance, depicted in  FIG.\ref{eepipi-fig}b,
as in the context of the vector dominance
model.
We refer to this process as the two-step
reaction, whereas the former, more general one, is the one-step reaction.
The corresponding transition amplitude then takes
the form
\begin{eqnarray}
T_2=\langle\pi^+\pi^-| V_{\overline qq}
|\rho\rangle\langle\rho|G|\rho\rangle\langle\rho|\overline
qq\rangle\langle\overline qq|T|e^+e^-\rangle
\end{eqnarray}
where $\langle\rho|\overline qq\rangle$ is simply the wave
function of the intermediate meson $\rho$,
$\langle\rho|G|\rho\rangle$ the Green function describing the
propagation of the intermediate meson, and
$\langle\pi^+\pi^-|V_{\overline qq}|\rho\rangle$ the transition
amplitude of $\rho^0$ annihilation into a $\pi^+\pi^-$ pair.
\par
The size parameter of the $\rho$ meson associated with its wave
function may be determined by studying the reaction
$\rho^0\rightarrow e^+e^-$. The transition amplitude of a vector
meson annihilation into an electron-positron pair takes the
general form
\begin{eqnarray}\label{rhotoee}
T=\langle e^+e^-|T|q\overline q\rangle\langle q\overline q|V\rangle
\end{eqnarray}
where $|V\rangle$ is the vector meson state (see Appendix A), and
$\langle e^+e^-|T|q\overline q\rangle$ the transition amplitude of
a quark-antiquark pair to an electron-positron pair. The
transition amplitude can be evaluated by a standard method as for
example outlined in \cite{hadrontransitions}. One has
\begin{eqnarray}\label{eetoqq}
\langle e^+e^-|T|q\overline q\rangle  = -  \frac{e_q \, e}{s}
\,\overline u_e(p_{e^-},m_{e^-})\gamma^\mu\,v_e(p_{e^+},m_{e^+})
\,\overline v_q(p_{\overline q},m_{\overline
q})\gamma_\mu\,u_q(p_q,m_{q})
\end{eqnarray}
where $s=(p_q+p_{\overline q})^2$, $e_q$ is the charge of quarks,
and the Dirac spinors are normalized according to $\overline
uu=\overline vv=2m$. In the small quark momentum approximation,
the decay width for the transition of a vector meson to an electron-positron pair can
be easily evaluated. One has
\begin{eqnarray}\label{cs}
\Gamma_{\rho^0\rightarrow e^+e^-} =
\frac{16\pi\alpha^2Q^2}{M_V^2}|\psi(0)|^2
\end{eqnarray}
where $Q^2$ is the squared sum of the charges of the quarks in the
meson, with $Q^2=1/2$ for $\rho$, 1/18 for $\omega$ and 1/9 for
$\phi$, and $\psi(0) = 1/(\pi b^2)^{3/4}$ is the coordinate space wave function
of the vector meson at the origin.
Using as an input $M_\rho=0.7758$ GeV, $\alpha=1/137$
and the experimental value of
$\Gamma_{\rho^0\rightarrow e^+e^-}=7.02\pm 0.11$ KeV, we get
$b=3.847$ GeV$^{-1}$
for the size parameter of the $\rho$ meson with the spatial wave
function set up in the harmonic oscillator
approximation (see details in Appendix A).
The size parameter $b$ in Eq.~(\ref{mspatial})
might be slightly different from meson to meson.

We use the reaction $\rho^0\rightarrow\pi^+\pi^-$ to determine the
effective strength parameter $\lambda$ in the quark-antiquark
$^3P_0$ vertex
\begin{eqnarray}\label{vertexij}
V_{ij} = \lambda\,\vec\sigma_{ij}\cdot(\vec p_{i}-\vec
p_{j})\,\hat F_{ij}\,\hat C_{ij}\,\delta(\vec p_{i}+\vec p_{j}) =
\lambda\sum_{\mu}\sqrt{\frac{4\pi}{3}}
(-1)^\mu\sigma^\mu_{ij}y_{1\mu}(\vec p_{i}-\vec
p_{j})\,\hat{F_{ij}}\,\hat{C_{ij}} \,\delta(\vec p_{i}+\vec p_{j})
\end{eqnarray}
where $y_{1\mu}(\vec q\, )=| \, \vec q \, | \, Y_{1\mu}(\hat q)$,
$\vec\sigma_{ij}=(\vec\sigma_{i}+\vec\sigma_{j})/2$, $\vec p_i$
and $\vec p_j$ are the momenta of quark and antiquark created out of the
vacuum. $\hat F_{ij}$ and $\hat C_{ij}$ are the
flavor and color operators projecting a quark-antiquark pair to the respective
vacuum quantum numbers. The derivation and interpretation of
the quark-antiquark $^3P_0$ dynamics may be found
in literature~\cite{3P0,tueb1}.

The decay width of the reaction $\rho^0\rightarrow\pi^+\pi^-$
takes the form
\begin{eqnarray}
\Gamma_{\rho^0\rightarrow\pi^+\pi^-} \, = \,
\frac{\pi}{4} \, M_\rho^2 \, \sqrt{1 -  \frac{4  M_\pi^2}{M_\rho^2}} \,
|T_{\rho^0\rightarrow\pi^+\pi^-}|^2,
\end{eqnarray}
where $T_{\rho^0\rightarrow\pi^+\pi^-}$ is the corresponding
transition amplitude defined in the center-of-mass system.
Substituting $T_{\rho^0\rightarrow\pi^+\pi^-}$ as calculated in our
approach (see Appendix A) we get
\begin{eqnarray}
\Gamma _{\rho^0\rightarrow\pi^+\pi^-} &=& \lambda^2\,\left(\frac{2}{3}\right)^7
\sqrt{\pi}M_\rho\,(bk)^3\,e^{-\frac{1}{6}\,b^2k^2}
\end{eqnarray}
where $M_\rho$ is the mass of $\rho$ meson and
$k = \sqrt{M_\rho^2/4 - M_\pi^2}$ the momentum of
the outgoing pions. The result obtained here is consistent with the ones
of Refs.~\cite{3P0,kokoski,barnes}, the different magnitude of the
strength parameter $\lambda$ just depends on the different
normalization of the $^3P_0$ vertex. With the size parameter
$b=3.847$ GeV$^{-1}$ determined from
the reaction $\rho^0\rightarrow e^+e^-$,
the experimental value $\Gamma=150$ MeV for the decay width of
$\rho^0\rightarrow\pi^+\pi^-$ requires the effective strength
parameter $\lambda$ to take the value $\lambda = 0.98\,.$
In~\cite{kokoski}, the size parameter $b$ is taken to be
2.5 GeV$^{-1}$ and the effective strength is fitted to be 0.39,
which according to our normalization corresponds to $\lambda=0.96$.
\par
Based on the evaluations for the reactions
$\rho^0\rightarrow\pi^+\pi^-$ and $\rho^0\rightarrow e^+e^-$, it
is straightforward to work out the transition amplitude of the two
step diagram shown in FIG.\ref{eepipi-fig}b in the reaction
$e^+e^- \rightarrow \pi^+\pi^-$
\begin{eqnarray}
T_{e^+e^- \rightarrow \pi^+\pi^-}&=&
T_{\rho^0\rightarrow\pi^+\pi^-}\frac{1}{E-M_\rho}
T_{e^+e^-\rightarrow\rho^0}.
\end{eqnarray}
The transition amplitude for the process
$\rho\rightarrow e^+e^-$ is
\begin{eqnarray}\label{rhotoee1}
T_{\rho\rightarrow e^+e^-}=  \langle e^+e^-|T|q\overline
q\rangle\langle q\overline q|V\rangle
=\int \frac{d\vec p_q\, d\vec p_{\overline q}}{(2\pi)^{3/2}2E_q}\,
\delta(\vec p_q+\vec p_{\overline q})\,
\psi_\rho(\vec p_q,\vec p_{\overline q})\,T_{q\overline
q\rightarrow e^+e^-}(\vec p_q,\vec p_{\overline q})
\end{eqnarray}
where $\psi_\rho$ is the wave function of the $\rho$ meson in momentum
space, and $T_{q\overline q\rightarrow e^+e^-}$ is given in
Eq.~(\ref{eetoqq}). The delta function $\delta(\vec p_q+\vec
p_{\overline q})$ indicates that we work in the $\rho$ meson rest
frame.

Note that only the $P$-wave contributes to the process $e^+e^- \to
\pi^+\pi^-$ since the spin of the intermediate $\rho$ is 1.
Furthermore, there is no free parameter since the size and the
effective strength parameters have been determined by the processes
$\rho^0\rightarrow e^+e^-$ and $\rho^0\rightarrow\pi^+\pi^-$,
respectively.
\begin{figure}
\centerline{\hspace{-0.5in}
\epsfig{file=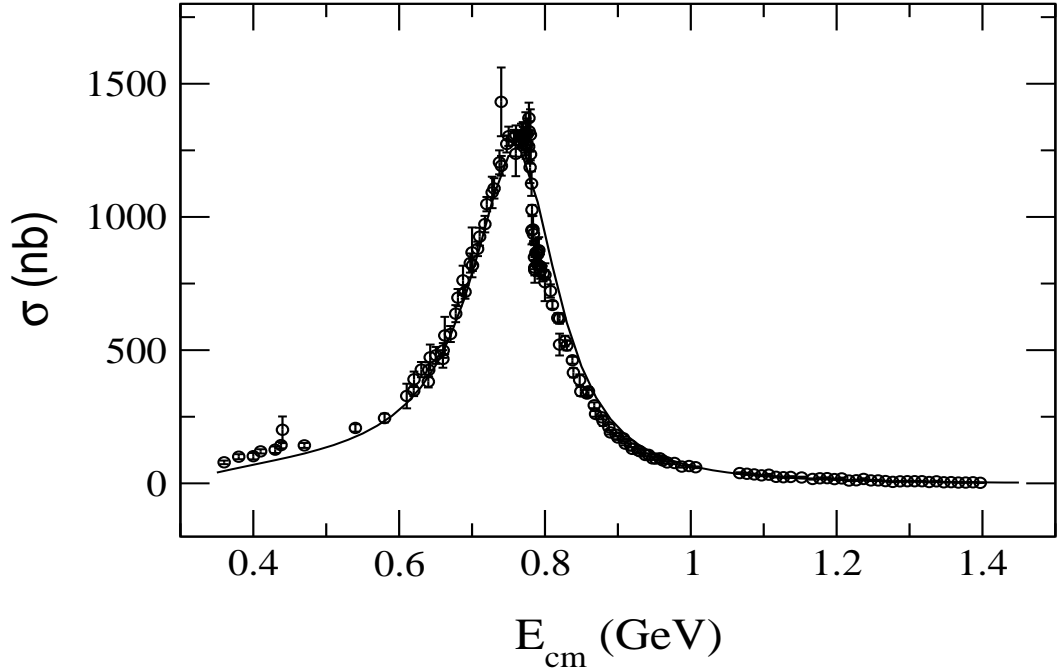,width=5.5in,height=3.5in,angle=0}}
\vspace{0.2in} \caption{Theoretical prediction (solid line) in the
two-step model shown in FIG.\ref{eepipi-fig}b for the cross
section of the reaction $e^+e^-\rightarrow\pi^+\pi^-$ compared
with experimental data taken
from~\cite{data-eepipi1,data-eepipi2}} \label{eepipi-sig}
\end{figure}

In FIG.\ref{eepipi-sig} we give the prediction for the cross
section of the reaction $e^+e^-\rightarrow\pi^+\pi^-$ in the model
for the two step process. The result seems to indicate that the
reaction $e^+e^-\rightarrow\pi^+\pi^-$ is completely dominated by
the intermediate vector meson.
One may therefore conclude that the one step process is completely
saturated by the relevant resonances entering at this energy scale.

\section{Reaction $e^+e^- \rightarrow \overline N N$}

In the following we assume that the reaction
$e^+e^-\rightarrow\overline NN$ is described by a two-step
process, just as for the reaction $e^+e^-\rightarrow\pi^+\pi^-$.
This is again consistent with the vector dominance model.
\par
\begin{figure}
\centerline{\hspace{-0.5in}
\epsfig{file=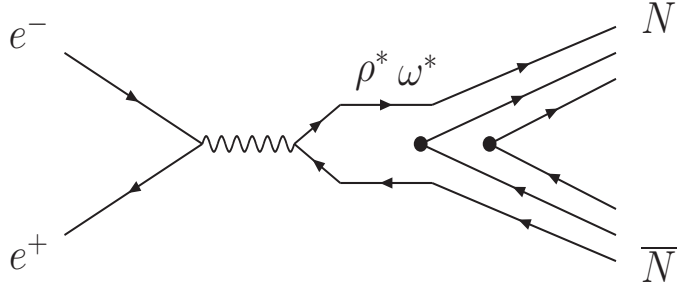,width=3.5in,height=1.5in,angle=0}}
\vspace{0.2in} \caption{Electron-positron annihilation into
nucleon-antinucleon pairs in a two-step process via intermediate
vector meson states.} \label{eeNN-fig}
\end{figure}

Here we study the two-step process shown in FIG.\ref{eeNN-fig}:
the $e^+e^-$ pair annihilates into a virtual time-like photon,
the photon decays into a $\overline qq$ pair, the $\overline qq$
pair forms a virtual vector meson, finally the virtual vector
meson is dressed by two additional quark-antiquark pairs created
out of the vacuum to form a baryon pair. The meson $\rho(2150)$
with the quantum number $I^G(J^{PC})=1^+(1^{--})$ is a good
candidate~\cite{particletable} for such an intermediate state. The
transition amplitude in such a two step process takes the form
\begin{eqnarray}\label{Ttwo}
T_{e^+e^- \rightarrow \overline N N} =
\langle \overline NN| V(^3P_0)|V\rangle\langle
V|G|V\rangle\langle V|\overline qq\rangle\langle\overline
qq|T|e^+e^-\rangle
\end{eqnarray}
Here $\langle V|\overline qq\rangle$ is simply the wave function
of the intermediate vector meson with both isospin $I=0$ and 1
(see Appendix B), $\langle V|G|V\rangle$ the Green's function
describing the propagation of the intermediate vector meson,
$\langle\overline NN|V(^3P_0)|V\rangle$ the transition amplitude
of the intermediate meson annihilation into a nucleon-antinucleon
pair, and $\langle\overline qq|T|e^+e^-\rangle$ the transition
amplitude of an electron-positron pair to a quark-antiquark pair
as given in Eq.~(\ref{eetoqq}). The transition amplitude $\langle
V|\overline qq\rangle\langle\overline qq|T|e^+e^-\rangle$ for the
process of the intermediate vector meson to an election-position
pair is defined as in Eq.~(\ref{rhotoee}) but with different meson
wave functions. The evaluation of the transition amplitude
$\langle\overline NN|V(^3P_0)|V\rangle$ is worked out
in Appendix B.
The energy scale of the intermediate $q^2\bar q^2$ state in the
$V\to N\bar N$ transition is simply set by associating in average
an equal share of the total energy to each valence quark involved.
\par
Considering that both isospin $I=0$ and 1 vector
mesons could be the intermediate states for the reaction
$e^+e^-\rightarrow\overline NN$, we have the transition amplitudes
\begin{eqnarray}
T_{e^+e^-\rightarrow\overline pp} &=&
\frac{1}{\sqrt{2}}[T(e^+e^-\rightarrow V(I=1)\rightarrow\overline
NN)+T(e^+e^-\rightarrow V(I=0)\rightarrow\overline NN)] \nonumber
\\
T_{e^+e^-\rightarrow\overline nn} &=&
\frac{1}{\sqrt{2}}[T(e^+e^-\rightarrow V(I=1)\rightarrow\overline
NN)-T(e^+e^-\rightarrow V(I=0)\rightarrow\overline NN)]
\end{eqnarray}
for the reactions $e^+e^-\rightarrow\overline pp $  and
$e^+e^-\rightarrow\overline nn$, respectively. It is clear that the
cross sections of the reactions $e^+e^-\rightarrow\overline pp$
and $e^+e^-\rightarrow\overline nn$ would be the same if either a
single isospin 0 or isospin 1 vector meson dominates the
intermediate state at this energy scale. However, the experimental
ratio of Eq.~(\ref{ratio}) indicates that at least two vector
mesons with isospin 0 and 1 are involved as intermediate states.
In addition to the confirmed vector meson $\rho(2150)$, there are
clues~\cite{omega1,omega2} for the existence of an $\omega$-like
meson lying in the energy region near the $\overline NN$
threshold. The vector meson with isospin $0$ has mass and width
of about 2150 MeV and 220 MeV, respectively. The contribution of
the $\omega(2150)$ as well as the $\rho(2150)$ are included in our
calculation. The mesons $\rho(2150)$ and $\omega(2150)$ are
assumed to be superpositions of $3S$ and $2D$ states, considering
that the lower lying states have been occupied ( see Table I).
\begin{figure}
\centerline{\hspace{-0.5in}
\epsfig{file=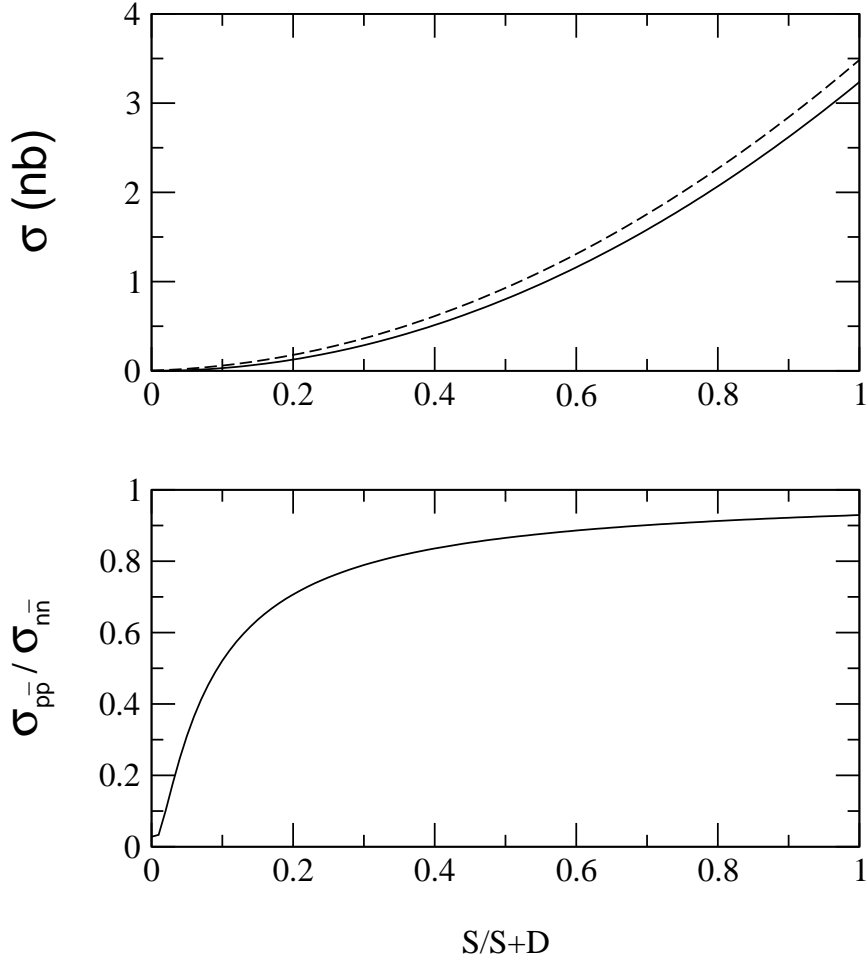,width=4.5in,height=5.in,angle=0}}
\vspace{0.2in} \caption{Model predictions for the cross sections
of the reactions $e^-e^+\rightarrow\overline nn$ (dashed line in
the upper figure) and $e^-e^+\rightarrow\overline pp$ (solid line
in the upper figure) and for the ratio
$\sigma(e^+e^-\rightarrow\overline
pp)/\sigma(e^+e^-\rightarrow\overline nn)$ (in the lower figure)
versus the S-wave probability of the meson $\rho(2150)$.}
\label{eeNN-sig}
\end{figure}
\begin{table}
\begin{center}
\label{rhoomega}
\caption{$\rho$ and $\omega$ mesons coming in
pairs}
\vspace*{.3cm}
\begin{tabular}{|c|c|c|}
\hline
1S & $\rho(770)$ & $\omega(782)$  \\
\hline
2S & $\rho(1450)$ & $\omega(1420)$ \\
\hline
1D & $\rho(1700)$ & $\omega(1650)$   \\
\hline
3S or 2D & $\rho(2150)$ & $\omega(2150)$ ? \\
\hline
\end{tabular}
\end{center}
\end{table}

It is found in our study that the experimental data suggest a
$\omega(2150)$ being in a $D$ wave, and prefer the $\rho(2150)$
meson as a mixture of $S$ and $D$ waves. It may be interesting to
mention that the work~\cite{barnes}, which studied the decay of
higher quarkonia in the $^3P_0$ quark model, reveals that the
lower energy counterparts $\rho(1450)$ and $\rho(1700)$ of the
meson $\rho(2150)$ are likely to be mixtures of $2S$ or $1D$
states.

Presented in FIG.\ref{eeNN-sig} are the predictions of the present
model for the total cross sections of the reactions
$e^+e^-\rightarrow\overline pp$ and $e^+e^-\rightarrow\overline
nn$ at the $\overline NN$ threshold, with the $\omega(2150)$ in
the $2D$ state and the $\rho(2150)$ varying from the $2D$ state to
the $3S$ state. Note that, except the parameters describing the
$S$ and $D$ wave admixture in the mesons $\rho(2150)$ and
$\omega(2150)$, there are no more free parameters. For the size
parameter we employ $b=3.847$ GeV$^{-1}$ as already fixed in the
reaction $\rho^0\rightarrow e^+e^-$, the $^3P_0$ strength with
$\lambda=0.98$ is fixed in the reaction
$\rho^0\rightarrow\pi^+\pi^-$. The size parameter of the nucleon
with $a=3.1$ GeV$^{-1}$ is fixed from other
considerations~\cite{asize,tueb1}, and the masses and widths of
the mesons $\rho(2150)$ and $\omega(2150)$ are taken
from~\cite{particletable,omega1,omega2}. The energy denominator of
the intermediate $q^2\bar q^2$ state $\Delta E$ is roughly
approximated as ${\Delta E} = E_{c.m.}/3\,,$ assuming that the
reaction energy $E_{c.m.}$ is shared by the six quarks
equally~\cite{thomas,yankaon}.

With $\omega(2150)$ in the $2D$ state and $\rho(2150)$
half in the $3S$ and half in the $2D$ state, we get
\begin{eqnarray}\label{section}
\sigma(e^+e^-\rightarrow\overline pp) \,
\approx \,  0.65 \,\,\mbox{nb}\,, \hspace*{1cm}
\sigma(e^+e^-\rightarrow\overline nn)\, \approx \,  0.76 \,\,\mbox{nb}
\end{eqnarray}
and hence
\begin{eqnarray}
\frac{\sigma(e^+e^-\rightarrow\overline
pp)}{\sigma(e^+e^-\rightarrow\overline nn)}&\approx& 0.85
\end{eqnarray}
The model results for the cross sections of Eq.~(\ref{section})
are sensitive to the length parameters $b$ and $a$ and the
effective parameter $\lambda$. However, the ratio
$\sigma(e^+e^-\rightarrow\overline
pp)/\sigma(e^+e^-\rightarrow\overline nn)$ is of course
independent of the strength parameter and rather independent of
the length parameters involved.

\section{Conclusions}
The puzzling experimental result that
$\sigma(e^+e^-\rightarrow\overline
pp)/\sigma(e^+e^-\rightarrow\overline nn) < 1$ can be understood
in the framework of a phenomenological nonrelativistic quark
model. All parameters employed in the model, except the ones
describing the mixture of the $S$ and $D$ waves for the
intermediate vector mesons $\rho(2150)$ and $\omega(2150)$, are
not free but determined by other reactions.

The experimental data suggest the existence of a $D$-wave $\omega$
meson with a mass of about 2100 MeV. The conclusion is quite
general, independent of the special values of the size parameters
$a$, $b$ and the $^3P_0$ strength $\lambda$.

\newpage

{\bf Acknowledgments}

\vspace*{.5cm}

\noindent This work was supported in part by the National
Research Council of Thailand (NRCT) under Grant No.~1.CH7/2545 and
the National Natural Science Foundation of China under Grant
Nos.~19975074 and 10275096, by the DFG under contracts FA67/25-3 and GRK683.
This research is also part of the EU Integrated Infrastructure Initiative
Hadronphysics project under contract number RII3-CT-2004-506078 and
President grant of Russia "Scientific Schools"  No. 1743.2003.

\appendix
\section{Transition $\rho\rightarrow\pi^+\pi^-$ in $^3P_0$ model}
\begin{figure}
\centerline{\hspace{-0.5in}
\epsfig{file=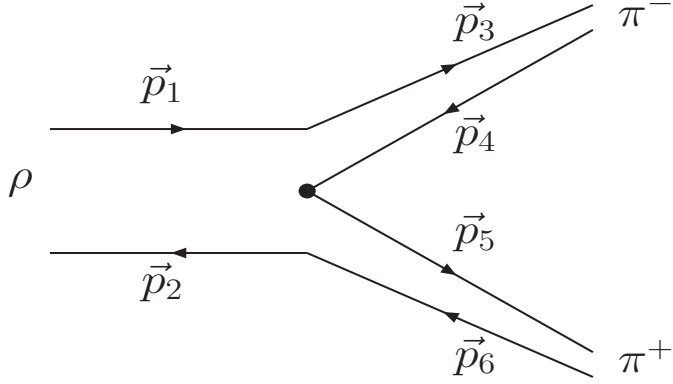,width=3.5in,height=2.0in,angle=0}}
\vspace{0.2in} \caption{$\rho\rightarrow\pi^+\pi^-$ in the $^3P_0$
nonrelativistic quark model.} \label{rhopipi}
\end{figure}

We study the reaction $\rho\rightarrow\pi^+\pi^-$ shown in
FIG.\ref{rhopipi} to determine the effective strength parameter
$\lambda$ in the quark-antiquark $^3P_0$ vertex of
Eq.~(\ref{vertexij}). The $\vec\sigma_{ij}$ in the
vertex can be understood as a operater acting on a quark and
antiquark state, or it projects a quark-antiquark pair onto a
spin-1 state. It can be easily proven that
\begin{eqnarray}
\langle
0,0|\sigma^\mu_{ij}|[\bar{\chi_i}\otimes\chi_j]_{JM}\rangle &=&
(-1)^M\sqrt{2}\delta_{J,1}\delta_{M,-\mu}.
\end{eqnarray}
Concerning SU(2) flavor a quark-antiquark pair which annihilates into
the vacuum must have zero isospin. So the operator $\hat F_{ij}$
has the similar property
$\langle 0,0|\hat F_{ij}|T,T_z\rangle
=\sqrt{2}\delta_{T,0}\delta_{T_z,0}\,. $
For the color part, one simplify has
$\langle 0,0|\hat C_{ij}|q^i_\alpha\overline q^j_\beta\rangle
=\delta_{\alpha\beta}$
where $\alpha$ and $\beta$ are color indices.
The transition amplitude for meson decay
into two mesons in the $^3P_0$ model shown in FIG.\ref{rhopipi}
is defined as $T=\langle \Psi_i|V_{45}^\dagger|\Psi_f\rangle$,
where $|\Psi_i\rangle$ and $|\Psi_f\rangle$ are the
initial and final states, respectively. For simplicity, we consider here only
the $S$-wave mesons, that is, all the mesons involved have orbital
angular momentum equal to 0. The initial state is simply the one
meson wave function (WF) having the form
\begin{eqnarray}\label{mspatial}
|\Psi_i\rangle=N\, e^{-\frac{1}{8}\,b^2(\vec p_1-\vec
p_2)^2}\left[\frac{1}{2}^{(1)}\otimes\frac{1}{2}^{(2)}\right]_{S_i,M_i}
\left[\frac{1}{2}^{(1)}\otimes\frac{1}{2}^{(2)}\right]_{T,T_z}
\end{eqnarray}
We have spin $S_i=1$ and isospin $T_i=1$ for the $\rho$
meson, and the isospin projection $T_z=0$ for $\rho^0$. Here we
have employed the harmonic oscillator interaction between quark
and antiquark.
The final state $|\Psi_f\rangle$ is formed by coupling the WFs of the
two final mesons. For two $S$-wave mesons we have
\begin{eqnarray}
|\Psi_f\rangle &=&  N_1N_2\, e^{-\frac{1}{8}\,b^2(\vec
p_3-\vec p_4)^2} \, e^{-\frac{1}{8}\,b^2(\vec p_5-\vec p_6)^2}
\, \left[\left[\frac{1}{2}^{(3)}\otimes\frac{1}{2}^{(4)}\right]_{S_1}
\otimes\left[\frac{1}{2}^{(5)}\otimes\frac{1}{2}^{(6)}\right]_{S_2}
\right]_{S_f,M_f} \nonumber\\
&\times&
\left[\left[\frac{1}{2}^{(3)}\otimes\frac{1}{2}^{(4)}\right]_{T_1}
\otimes\left[\frac{1}{2}^{(5)}\otimes\frac{1}{2}^{(6)}\right]_{T_2}
\right]_{T,T_z}
\end{eqnarray}
The transition amplitude is derived as
\begin{eqnarray}
T_{spatial}&=& \lambda\,
\frac{2^{4}}{3^3\sqrt{3}\,\pi^{1/4}}b^{3/2}k\,e^{-\frac{1}{12}b^2k^2}
\end{eqnarray}
Note that we have, for simplicity, set the $\rho$ and $\pi$ mesons to have
the same size parameter $b$, that is
$N=N_1=N_2=(b^2/\pi)^{3/4}\,.$

\section{Transition $V \to \overline NN$ in $^3P_0$ model}
In the $^3P_0$ model the transition amplitude for a
vector meson decaying into a $\overline NN$ pair might be written
in the form
\begin{eqnarray}
\langle\overline NN|V(^3P_0)|V\rangle &\equiv& \langle\overline
NN|V_{25}^\dagger\frac{1}{\Delta E}V_{36}^\dagger|V\rangle \nonumber \\
&=& \frac{4\pi}{3}\,\lambda^2\,\frac{1}{\Delta E}\sum_{S'_z}
C(L\,J_z-S'_z,S'S'_z,1J_z)\cdot T_{color}T_{sf}T_{spatial}
\end{eqnarray}
where the Clebsch-Gordon coefficient $C(L\,J_z-S'_z,S'S'_z,1J_z)$
results from the spin-orbital coupling of the intermediate meson
having the orbital angular momentum $L=0,2$ and spin $S'$, and the
factor $1/\Delta E$ accounts for the energy propagation between
the two quark-antiquark vertices $V_{25}^\dagger$ and
$V_{36}^\dagger$ which are defined as in Eq.~(\ref{vertexij}).
Here we have supposed that $\Delta E$ is constant for a given
reaction energy as, for examplem, discussed in~\cite{yankaon,thomas}.
Using the
wave functions defined in the previous sections we get for
the color part $T_{color} = \frac{1}{\sqrt{3}}\,,$
for the spin-flavor part
\begin{eqnarray}
T_{sf} &=& \frac{1}{2} \,
\left\langle J^{[ij]} \right|_{S'S'_z}
(-1)^{\mu}\sigma^{25}_{-\mu}\hat F_{25}(-1)^{\nu}\sigma^{36}_{-\nu}
\hat F_{36} \sum_{J_{23},J_{56}} \,
\left| J^{[231;564]}\right\rangle_{SS_z}^{\rm Spin} \,
\left| J^{[231;564]} \right\rangle_{TT_z}^{\rm Flavor} \\
J^{[ij]} &=& \frac{1}{2}^{(7)} \otimes \frac{1}{2}^{(8)}, \,
J^{[ijk;lmn]} =
\left[\left(\frac{1}{2}^{(i)}\otimes\frac{1}{2}^{(j)}\right)_{J_{ij}}
\otimes\,\frac{1}{2}^{(k)}\right]_{1/2}\otimes
\left[\left(\frac{1}{2}^{(l)}\otimes\frac{1}{2}^{(m)}\right)_{J_{lm}}
\otimes\,\frac{1}{2}^{(n)}\right]_{1/2}\nonumber
\end{eqnarray}
and for the spatial part
\begin{eqnarray}
T_{spatial} &=& \int\prod d^3q_i\,\Psi^\dagger_{N\overline N}
Y^*_{1\mu}(\vec q_{25})\delta^{(3)}(\vec q_{25})
Y^*_{1\nu}(\vec q_{36})\delta^{(3)}(\vec q_{36}) \\
&\times&\Psi_{m}(\vec q_{78}) \delta^{(3)}(\vec q_{17})
\delta^{(3)}(\vec q_{48})\delta^{(3)}(\vec q_{123}-\vec k)
\delta^{(3)}(\vec q_{456}+\vec k)\nonumber
\end{eqnarray}
where $\vec q_{ij} = \vec q_{i} + \vec q_{j}$,
$\vec q_{ijk} = \vec q_{i} + \vec q_{j} + \vec q_{k}$,
$\Psi_{N\overline N}$ is the spatial wave function of the
$N\overline N$ state
\begin{eqnarray}
\Psi_{N\overline N}=N_b^2 \,
e^{-\frac{1}{4}\,a^2\,[ \, \vec q_{23}^{\, 2} \,
+ \, \vec q_{56}^{\, 2} \, ] }
\,
e^{-\frac{1}{12}\,a^2\, [ \, (\vec q_{12} - \vec q_{13})^2 \, + \,
(\vec q_{46} - \vec q_{45})^2 \, ] }
\end{eqnarray}
and $\Psi_m$ with $m=s, d$ the spatial wave function of
the intermediate meson which are taken as $3S$ and $2D$ states
\begin{eqnarray}
\Psi_{s}(\vec p\,)&=&
N_s \,e^{-\frac{1}{2}\,b^2p^2} \,
\left(\frac{15}{4}-5b^2p^2+b^4p^2\right)\,,\\
\Psi_{d}(\vec p\,)&=&N_{d}\,
e^{-\frac{1}{2}\,b^2p^2} \, (bp)^2 \,
\left(\frac{7}{2}-b^2p^2\right)
Y_{2L_z}(\hat p)\nonumber \,.
\end{eqnarray}
At the $\overline NN$ threshold, that is, $k\approx 0$, one may
evaluate $T_{spatial}$ analytically. For the process where the
vector meson is in a S-wave, we obtain
\begin{eqnarray}
T_{spatial} &=& 4\cdot(4\pi)\cdot
8\cdot\delta_{\mu,-\nu}(-1)^\nu\cdot N_sN_b^2
\left\{f(2,\alpha)\left[\frac{15}{4}f(4,\beta)-20b^2f(4,\beta)
+14b^4f(8,\beta)\right]
\right.
\nonumber\\
&-&\left.
f(4,\alpha)\left[\frac{15}{4}f(2,\beta)-20b^2f(4,\beta)
+14b^4f(6,\beta)\right]\right\}
\end{eqnarray}
For the process where the vector meson is in a D-wave,
we have
\begin{eqnarray}
T_{spatial} = 16\cdot(4\pi)\cdot 8 \cdot
b^2N_sN_b^2\cdot I\cdot
f(2,\alpha)\left[\frac{7}{2}f(6,\beta)-4b^2f(8,\beta)\right]
\end{eqnarray}
with
\begin{eqnarray}
I=\frac{3}{\sqrt{5}}\frac{1}{\sqrt{4\pi}}C(10,10,20)C(1\mu,1\nu,2\,\mu+\nu).
\end{eqnarray}
In the above equations $\alpha$ and $\beta$ are constants defined
as $\alpha=2a^2$ and $\beta=2b^2+6a^2\,.$
The function $f(n,u)$ is given as
\begin{eqnarray}
f(n,u)=\int_0^\infty\, dx\,x^n e^{-ux^2} \, = \, \frac{1}{2}
\, u^{-\frac{n+1}{2}}  \,
\Gamma\biggl(\frac{n+1}{2}\biggr) \,.
\end{eqnarray}


\begin{thebibliography}{20}
\bibitem{eebar1}
A.~Antonelli {\it et al.},
Nucl.\ Phys.\ B {\bf 517}, 3 (1998).
\bibitem{eebar2}
A.~Antonelli {\it et al.},
Phys.\ Lett.\ B {\bf 334}, 431 (1994);
D.~Bisello {\it et al.}  [DM2 Collaboration],
Z.\ Phys.\ C {\bf 48}, 23 (1990).
\bibitem{eebar3}
G.~Bardin {\it et al.},
Nucl.\ Phys.\ B {\bf 411}, 3 (1994).
\bibitem{Ellis}
J.~R.~Ellis and M.~Karliner,
New J.\ Phys.\  {\bf 4}, 18 (2002)
[arXiv:hep-ph/0108259].
\bibitem{3P0}
A.~Le Yaouanc, L.~Oliver, O.~Pene and J.~C.~Raynal,
Phys.\ Rev.\ D {\bf 8}, 2223 (1973);
Phys.\ Rev.\ D {\bf 9}, 1415 (1974);
Phys.\ Rev.\ D {\bf 11}, 1272 (1975).
\bibitem{tueb1}
M.~Maruyama, S.~Furui and A.~Faessler,
Nucl.\ Phys.\ A {\bf 472}, 643 (1987);
M.~Maruyama, S.~Furui, A.~Faessler and R.~Vinh Mau,
Nucl.\ Phys.\ A {\bf 473}, 649 (1987);
T.~Gutsche, M.~Maruyama and A.~Faessler,
Nucl.\ Phys.\ A {\bf 503}, 737 (1989).
\bibitem{dgmf}
C.~B.~Dover, T.~Gutsche, M.~Maruyama and A.~Faessler,
Prog.\ Part.\ Nucl.\ Phys.\  {\bf 29}, 87 (1992).
\bibitem {muhn}
A.~Muhm, T.~Gutsche, R.~Thierauf, Y.~Yan and A.~Faessler,
Nucl.\ Phys.\ A {\bf 598}, 285 (1996).
\bibitem{yanatom}
Y.~Yan, R.~Tegen, T.~Gutsche and A.~Faessler,
Phys.\ Rev.\ C {\bf 56}, 1596 (1997).
\bibitem{hadrontransitions}
A.~Le Yaouanc, L.~Oliver, O.~Pene and J.~C.~Raynal,
{\it Hadron Transitions in the Quark Model} (Gordon and Breach, Amsterdam, 1988).
\bibitem{kokoski}
R.~Kokoski and N.~Isgur,
Phys.\ Rev.\ D {\bf 35}, 907 (1987).
\bibitem{barnes}
E.~S.~Ackleh, T.~Barnes and E.~S.~Swanson,
Phys.\ Rev.\ D {\bf 54}, 6811 (1996)
[arXiv:hep-ph/9604355];
T.~Barnes, F.~E.~Close, P.~R.~Page and E.~S.~Swanson,
Phys.\ Rev.\ D {\bf 55}, 4157 (1997)
[arXiv:hep-ph/9609339].
\bibitem{data-eepipi1}
L.~M.~Barkov {\it et al.},
Nucl.\ Phys.\ B {\bf 256}, 365 (1985).
\bibitem{data-eepipi2}
R.~R.~Akhmetshin {\it et al.}  [CMD-2 Collaboration],
Phys.\ Lett.\ B {\bf 527}, 161 (2002)
[arXiv:hep-ex/0112031].
\bibitem{particletable}
S.~Eidelman {\it et al.}  [Particle Data Group Collaboration],
Phys.\ Lett.\ B {\bf 592}, 1 (2004).
\bibitem{thomas}
T.~Gutsche, R.~D.~Viollier and A.~Faessler,
Phys.\ Lett.\ B {\bf 331}, 8 (1994).
\bibitem{yankaon}
Y.~Yan, S.~W.~Huang and A.~Faessler,
Phys.\ Lett.\ B {\bf 354}, 24 (1995).
\bibitem{omega1}
A.~V.~Anisovich {\it et al.},
Phys.\ Lett.\ B {\bf 476}, 15 (2000).
\bibitem{omega2}
A.~V.~Anisovich {\it et al.},
Phys.\ Lett.\ B {\bf 507}, 23 (2001).
\bibitem{asize}
N.~Isgur and G.~Karl,
Phys.\ Rev.\ D {\bf 18}, 4187 (1978);
Phys.\ Rev.\ D {\bf 20}, 1191 (1979).
\end{thebibliography}
\end{document}